\documentclass[%
twocolumn,
 amsmath,amssymb,
 aps,
pra
]{revtex4-2}

\usepackage{graphicx}
\usepackage{dcolumn}
\usepackage{bm}
\usepackage{verbatim}
\usepackage{hyperref}
\usepackage{color,tabularx}

\newcolumntype{P}[1]{>{\centering\arraybackslash}p{#1}}

\usepackage{enumitem}
\usepackage{soul}

\begin{document}

\title{Finite-range bias in fitting three-body loss to the zero-range model}

\author{Sofia Agafonova, Mikhail Lemeshko, and Artem G. Volosniev}
\affiliation{Institute of Science and Technology Austria (ISTA),
    Am Campus 1, 3400 Klosterneuburg, Austria}
    
    \begin{abstract}

    We study the impact of finite-range physics on the zero-range-model analysis of three-body recombination in ultracold atoms.
We find that temperature dependence of the zero-range parameters can vary from one set of measurements to another as it may be driven by the distribution of error bars in the experiment, and not by the underlying three-body physics.
To study finite-temperature effects in three-body recombination beyond the zero-range physics, we introduce and examine a finite-range model based upon a hyperspherical formalism.
The systematic error discussed in the paper may provide a significant contribution to the error bars of measured three-body parameters.
\end{abstract}
    
\maketitle

Three-body recombination loss in cold-atom experiments provides an invaluable tool in fundamental studies of three-body physics, in particular of the Efimov effect~\cite{Efimov1970,Efimov1971,Jensen2011,Nielsen2001,Jensen2004,BRAATEN2006,Naidon2017,Greene2017,DIncao2018,Grimm2019}. Although many features of the experimental data are captured by zero-range models, current experiments also reveal finite-range effects~\cite{PhysRevA.91.063622,Wacker2018,Pires2014}, which require theoretical analysis of corresponding physics.

To analyze three-body loss, one considers the number of particles lost from the system per unit of time, $\alpha$. In ultracold dilute gases, $\alpha$  depends on a handful of quantities that characterize particle-particle interactions~\cite{BRAATEN2006,Naidon2017,Greene2017}. The first one is the scattering length $a$, which can be controlled using external fields~\cite{Chin2010}. Minimal zero-range models have two more parameters that define short-range  three-body physics and probability to recombine~\cite{BRAATEN2006}, denoted  (for $a<0$) as $a_-$ and $\eta$. Their values are obtained from experimental data~\cite{Naidon2017,Greene2017}, and often considered to be intrinsic to the few-body system at hand. However, it was observed that $a_{-}$ depends on temperature~\cite{PhysRevA.91.063622,Wacker2018}, contradicting theoretical expectations. This dependence attributed to the finite-range physics (always present in realistic systems) is modelled in our paper.  

We show that the temperature dependence of $a_-$ may be driven by the distribution of error bars in the experiment, and not by the underlying three-body physics. This is a consequence of the fact that the parameters $a_-$ and $\eta$ describe measurements only from the point of view of an incomplete (zero-range) theory -- they may contain (besides intrinsic few-body physics) information about the experiment. 

Our results add a possible systematic bias to the family of already known ones caused, for example, by high densities~\cite{PhysRevLett.123.233402} or uncertainties in the trap frequencies and atom number~\cite{Weber2003}. However, unlike the previously known issues with analysis of three-body recombination, one requires to deepen {\it theoretical} understanding of microscopic physics to mitigate the bias discussed below. 
This can be important for studies of the van der Waals universality, which provides an estimate of $a_-$ at zero temperature, $T=0$~\footnote{The value of $a_-$ is (nearly) universal for many [although not all~\cite{PhysRevLett.123.233402}] alkali atoms -- it is determined by the van der Waals length~\cite{Berninger2011,Wang2012,Schmidt2012,Naidon2014,Naidon2017,Greene2017,DIncao2018}.}.
 To test it in a laboratory, one performs a number of measurements at different temperatures, and extrapolates the fitted $a_-$ to the limit $T\to 0$~\cite{PhysRevA.91.063622,Wacker2018,PhysRevLett.123.233402}. 
As we shall demonstrate, this procedure may be inconclusive. It is an example of a much more general phenomenon --  ambiguity of fits based on universal theories in the presence of non-universal physics. The corresponding systematic errors are not well understood in the context of cold-atom set-ups. Although, they should be analyzed on a case-by-case basis, some physical intuition can be adopted from other branches of physics, in particular, from studies of a few nucleons~\cite{Furnstahl_2015,Carlsson2016}.

\begin{figure}
    \includegraphics[]{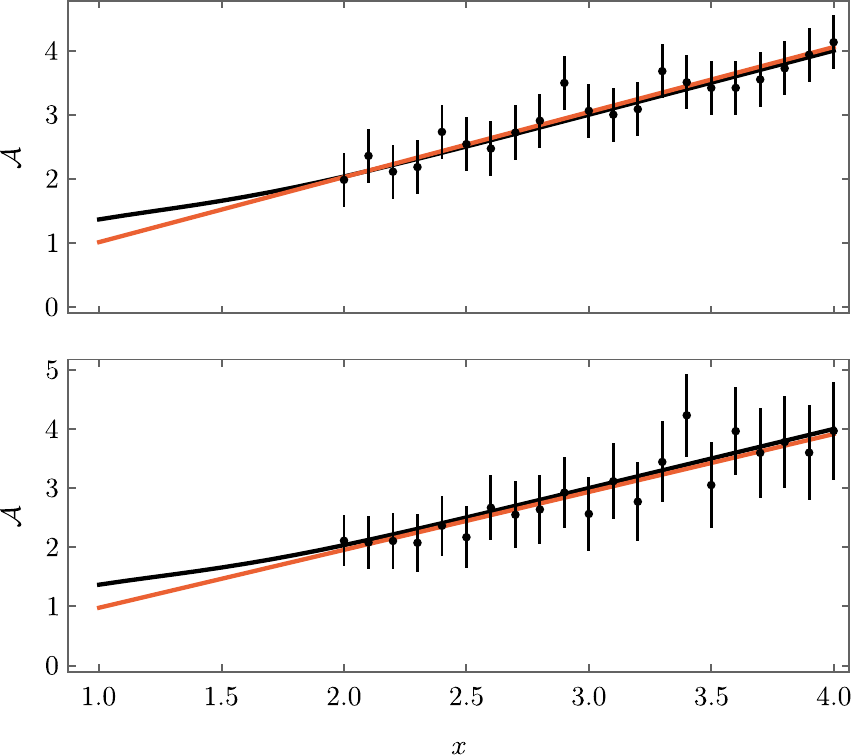}
    \caption{Illustration of the toy model for one representative set of parameters. The black curves show the exact model of Eq.~(\ref{eq:toy_mode_true_physics}). The dots with error bars correspond to generated `experimental' data for  $x=\{2,2.1,2.2,...,4\}$. The data are drawn from the normal distribution with the mean given by Eq.~(\ref{eq:toy_mode_true_physics}), and the standard deviation $\epsilon_i/2$. The upper panel is for $\epsilon_1=0.4$, the lower panel is for $\epsilon_{2}=0.4 \mathcal{A}(x)/\mathcal{A}(x_1)$. The orange curves show the best linear fit to the `experimental' data. }
    \label{fig:toy_model}
\end{figure}

{\it Illustrative Toy Model.} Before discussing three-body recombination, let us discuss an example that provides insight into the fact that if a fitting model does not describe every relevant aspect of the data (incomplete or underfit model), then its parameters may depend on characteristics of the experiment. Moreover, the values extracted from different experiments may not overlap within respective error bars, leading to a systematic bias in the analysis. To illustrate this rather general statement, we introduce and discuss a toy model. The model is linear by design, i.e., the fitting function is a linear function of the parameters. This will allow us to gain some analytical insight into the problem.

Consider an artificial physical process described by
\begin{equation}
    \mathcal{A}(x)=x+e^{-x^2} x \;,
    \label{eq:toy_mode_true_physics}
\end{equation}
where $x$ is some parameter, e.g., a dimensionless length scale; $  \mathcal{A}(x)$ is an observable.   
Interpretation of Eq.~(\ref{eq:toy_mode_true_physics}) is as follows: (i) for $x\to\infty$ the system obeys the `universal' physics ($\mathcal{A}\to x$), (ii) for $x\to0$, some `non-universal' physics is important, which, for simplicity, is parameterized here by $e^{-x^2} x$.

Let us assume that there are two experiments that measure $\mathcal{A}$ at different values of $x$. Each experiment produces a data set 
$\{\{\mathcal{A}_i(x_1),\epsilon_i(x_1)\},\{\mathcal{A}_i(x_2),\epsilon_i(x_2)\},...,\{\mathcal{A}_i(x_N),\epsilon_i(x_N)\}\}$. Here, the subscript denotes the experiment. $\epsilon_i$ denotes the corresponding error in the measurement of $\mathcal{A}$.
It is assumed that the value of $x$ is known exactly in each experiment so that there are no associated error bars. It is also assumed that both experiments measure at identical values of $x$.  As will become clear later,
these assumptions are not essential.

To simulate data measured in each experiment, we draw random values $\mathcal{A}_i$ for each $x$ from a normal distribution with the mean given by Eq.~(\ref{eq:toy_mode_true_physics}), and the standard deviation $\epsilon_i/2$.
The two experiments differ {\it only} in the values of $\epsilon_i$. In the first experiment, we assume $\epsilon_1=\epsilon$, and in the second one $\epsilon_2(x_k)=\mathcal{A}(x_k)\epsilon/\mathcal{A}(x_1)$. Both choices appear logical -- the first corresponds to a fixed error, the second corresponds to an error proportional to the value of $\mathcal{A}$.  

To analyze the data, we assume that the functional dependence of the universal physics is known, i.e., it is known that $\mathcal{A}_i(x_k)\simeq x_k$ in the limit $x_k\to\infty$. Therefore, we fit the data with $a_i x$, where $a_i$ is a fit parameter~\footnote{We have checked that one can use other simple expressions, for example, $a_i x + b_i$, without changing the conclusion.}, see Fig.~\ref{fig:toy_model}. It is clear that the value of $a_i$ depends on the number of data points, $N$, as well as on the range of $x$, i.e., on $x_1$ and $x_N$. For example, if there are `sufficiently' many data points in the  universal regime, i.e., with $x\gg 1$, then the mean values of $a_1$ and $a_2$ should approach $1$. Here, we are interested in the scenario in which the parameter $a_i$ contains some information about non-universal physics, which is a typical experimental situation.   To take this into account, we fix $x_1=2$ and $x_N=4$. This region is `almost' universal, as $e^{-x_1^2}\sim 0.018$, however still contains some information about the small-$x$ region.

To determine $a_i$, we minimize $\chi^2$ (``chi-squared'')~\cite{numericalrecipes,Young2012}
\begin{equation}
\chi^2_i=\sum_{k=1}^N\left(\frac{\mathcal{A}_i(x_k)-a_i x_k}{\epsilon_i(x_k)}\right)^2.
\end{equation}
After differentiating $\chi^2_i$ with respect to $a_i$, we derive
\begin{equation}
    a_i=\frac{\sum_k [\mathcal{A}_i(x_k)x_k/\epsilon_i(x_k)^2]}{\sum_k [x_k/\epsilon_i(x_k)]^2 }.
\end{equation}
We assume that $x_{k+1}-x_k=\delta x\to 0$, i.e., the experiment has a fine grid in $x$. Furthermore, we assume that $\epsilon\to0$, i.e., measurements in both experiments enjoy tiny error bars. 
With these assumptions, we write
$a_1=\int_{x_1}^{x_N} \mathcal{A}(x)x\mathrm{d}x/\int_{x_1}^{x_N} x^2\mathrm{d}x$ and $ a_2=\int_{x_1}^{x_N} x\mathcal{A}^{-1}(x)\mathrm{d}x/\int_{x_1}^{x_N} (x\mathcal{A}^{-1}(x))^2\mathrm{d}x$.
We see that the first experiment leads to $a_1\simeq 1.0010$. The second experiment yields $a_2\simeq 1.0020$~\footnote{$a_i>1$ because $\mathcal{A}(x)/x>1$ (see Eq.~(\ref{eq:toy_mode_true_physics})).}. Even though the two values are very close to each other, they are different. This reflects the fact that the first experiment trusts all points equally ($\epsilon_1(x_1)=\epsilon_1(x_N)$), whereas, the second has more confidence in non-universal points (e.g., $\epsilon_2(x_1)/\epsilon_2(x_N)\simeq x_1/x_N<1$).
Note that within the realm of each numerical experiment, the values of $a_1$ and $a_2$ are exact as $\epsilon\to0$, and contradict each other. [It is easy to check numerically that $a_1\neq a_2$ even if we assume that there is some variation in the error bars, e.g., if $\epsilon$ is a random number drawn from a normal distribution with the mean $\tilde\epsilon$ and the standard deviation $\tilde \epsilon/10$.]

The systematic error discussed above is based on two facts. First, the universal model is an underfit model, i.e., it does not describe the data `sufficiently' well when $\epsilon\to0$. Second, we {\it systematically} force the fitting procedure to trust `non-universal' physics more in the second experiment. One can improve the fitting procedure in this section by introducing other terms, which mimic `non-universal' physics, to the fitting function.
Alternatively, if one has some knowledge of the second term in Eq.~(\ref{eq:toy_mode_true_physics}), one can set the upper bound on the error $\epsilon_i$~\cite{Carlsson2016} or use Bayesian parameter estimation~\cite{Wesolowski_2016}.
We refrain from utilizing these options here, as (i) there is limited understanding of finite-range range effects on three-body recombination at finite temperature (ii)
our aim is to mimic the state-of-the-art analysis of three-body recombination in cold atoms.

The toy model presented in this section is artificial and contains assumptions (e.g., $\epsilon\to0$) that might be hard to satisfy experimentally. In spite of this, it illustrates the fact that the parameter $a$ corresponds to a physical quantity only if $x_1\gg 1$, otherwise, $a$ depends on the experimental protocol even for very accurate and dense data sets. To illustrate how the corresponding error might enter the analysis of three-body recombination, we shall simulate the standard routine for analyzing experiments. To this end, we introduce a finite-range model to generate `experimental' data. Then we fit these data using the zero-range model of Refs.~\cite{Braaten2008,Rem2013}.

{\it Model to simulate three-body loss.} Typically, a three-boson problem is notoriously difficult to solve. However, in the limit of low energies and short-range interactions one can obtain an accurate solution with a single differential equation for the hyper-radial wave function, $f$, in the adiabatic approximation (for review, see~\cite{Nielsen2001})
\begin{equation}
\label{eqn:schroedinger}
    \begin{aligned}
    \left(-\frac{d^{2}}{d\rho^{2}}+\frac{\nu^{2}(\rho)-1/4}{\rho^{2}}-\frac{2mE}{\hbar^{2}}\right)f\left(\rho\right)=0,
    \end{aligned}
\end{equation}
where $E$ is the energy, $m$ is the mass of a boson, and $\rho$ is the hyper-radius~\footnote{  $\rho=\sqrt{2/3}\sqrt{r_{1}^{2}+r_{2}^{2}+r_{3}^{2}-\boldsymbol{r_{1}\cdot r_{2}}-\boldsymbol{r_{2}\cdot r_{3}}-\boldsymbol{r_{1}\cdot r_{3}}}$, $\boldsymbol{r_{i}}$ - coordinate of $i$th particle}. The function $\nu(\rho)$ determines the effective three-body potential from two-body interactions. In spite of simplicity of Eq.~(\ref{eqn:schroedinger}), it provides a valuable tool in studies of universal properties of three-body states~\cite{Fedorov1993,Nielsen1998} and associated with them losses in cold gases, see, e.g.,~\cite{Jonsell2006,Sorensen2013}.

For a fixed value of $\rho$, the parameter $\nu$ solves the equation 
\begin{equation}
\label{eqn:potential}
    \frac{8}{\sqrt{3}}\sin{\frac{\nu \pi}{6}}-\nu\cos{\frac{\nu \pi}{2}}=\sqrt{2}\rho\left(\frac{1}{|a|}+F\right)\sin{\frac{\nu \pi}{2}},
\end{equation}
where $a$ is the scattering length, and $F$ contains information about finite-range corrections. 
If $F=0$, then Eq.~(\ref{eqn:potential}) leads to the `zero-range' model, see, e.g., Refs.~\cite{Braaten2008,Rem2013,Wacker2018}. It describes three-body recombination rate accurately, assuming that the fitting parameters $a_-$ and $\eta$ might depend on the temperature~\cite{PhysRevA.91.063622,Wacker2018}. The aim of this section is to provide an algorithm for generating `experimental' data of three-body recombination using a finite-range model with $F\neq0$~\footnote{It might be more appropriate to refer to `the finite-range model' as an `extended zero-range model'. For simplicity, we do not do it here.}. 

To investigate finite-range effects, we shall use the following expression of $F$~\cite{Fedorov_2001,Platter2009},
\begin{equation}
    F(\rho)=\frac{R}{4}\frac{\nu^2}{\rho^2},
    \label{eq:Frho}
\end{equation}
where $R$ is  a new length scale in the problem -- the effective range parameter that appears in two-body scattering.  It is worth noting that studies based upon hyperspherical formalism as well as effective field theories suggest the existence of one more three-body parameter once range corrections are considered~\cite{Thogersen2009,Ji_2010,JI2012}. As our aim here is to design a minimal  model for studying systematic errors possible in the experiment, we refrain from introducing this additional parameter.  We note that in the hyperspherical formalism, its effect is cancelled (at least partially) by non-adiabatic corrections~\cite{Thogersen2009}.

In one-channel atom-atom scattering, the parameter $R$ is typically positive, see, e.g.,~\cite{Gao1998,Flambaum1999}. However, in multi-channel problems, which are more suitable for modeling ultracold set-ups, this parameter is negative~\cite{Bruun2005}, thus, we shall assume $R<0$ (see~\cite{SM} for a discussion of the case with $R>0$)\footnote{Note that in cold-atom experiments, the parameter $R$ can also depend on the external magnetic field, and hence the scattering length~\cite{Sorensen2013a}. For simplicity, we do not consider this dependence here. This assumption is reasonable when the background scattering length is much smaller than the scattering length engineered in the experiment}.  

The solution to Eq.~(\ref{eqn:potential}) as a function of $\rho/|a|$ is plotted in~\cite{SM}. The main features of the solutions are as follows. In the zero-range model ($R=0$), $\nu^{2}\left(0\right)\approx -1$. This leads to a (super) attractive $-1.26/r^{2}$ potential in Eq.~(\ref{eqn:schroedinger}), which supports
an infinite number of bound states with the ground state of infinite negative energy – the Thomas collapse~\cite{Thomas1935} (for review, see~\cite{Nielsen2001,BRAATEN2006,Naidon2017}). The
collapse occurs only for $R=0$. In a finite-range model ($R<0$), the solution to Eq.~(\ref{eqn:potential}) in the limit $\rho\to0$ is determined by $R$; $\nu^2(0)$ vanishes  and the Thomas collapse does not occur. The long-range part is determined mainly by the scattering length, see~\cite{SM}.

In general if $\rho\gg \sqrt{|R a|}$, then $\nu^2$ is given approximately by $\nu_{ZR}^2$ -- the solution of Eq.~(\ref{eqn:potential}) with $F=0$. [This estimate is obtained by comparing $1/|a|$ and $F$ assuming that $\nu$ is of the order of unity.]. 
In this limit we can derive
\begin{equation}
    \nu^2\simeq \nu^2_{ZR}+\frac{R}{\rho}g\left(\frac{\rho}{a}\right),
    \label{eq:large_R_limit}
\end{equation}
where $g$ is given in~\cite{SM}. In our numerical simulations, this expansion is accurate already for $\rho \gtrsim 2 |R|$. Note that the parameter $R$ enters linearly in this expression. Therefore, positive (negative) values of $R$ lead to larger (smaller) values of the effective potential.

\begin{figure}[t]
    \includegraphics{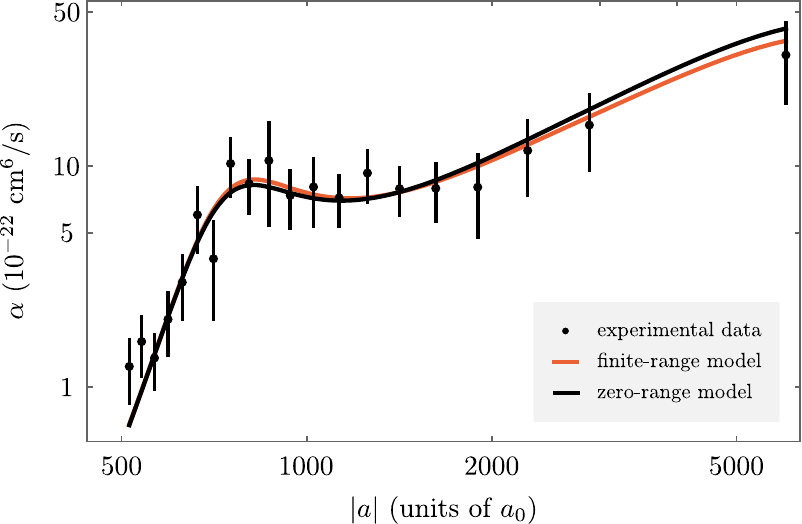}
    \caption{Recombination coefficient, $\alpha$, from the experiment of Ref.~\cite{Wacker2018} at $T=$178 nK (dots with error bars). The figure also shows the fit to the zero-range model (black) and to the finite-range model (orange) with $R=-55a_{0}$ ($a_0$ is the Bohr radius). The value of $|R|$ is chosen close to the corresponding van der Waals length $R_{vdW}=64.53 a_0$~\cite{JIANG2015158}.}
    \label{fig:datafit}
\end{figure}

\begin{table}
\begin{tabular}{|P{1.1cm} P{1.1cm} P{1.1cm} P{0.9cm} P{0.9cm} P{0.9cm} P{0.9cm}|} 
\hline
$T$ (nK) & $|a_{-}|$ & $\eta$ & $\chi_{0}^{2}$ & $r$ & $\phi$ &$\chi^{2}$ \\ 
\hline\hline
178 & 772 & 0.24 &  0.5 & 1.7 & 2.2 & 0.4  \\
\hline
192 & 718 & 0.22 & 0.5  & 1.3 & 2.1 & 0.4  \\
\hline
286 & 824 & 0.25 &0.2 & 2.1 & 2.2 & 0.1 \\
\hline
304 & 769 & 0.31 & 0.4 & 1.9 & 2.0 & 0.3 \\
\hline
\hline
avg. & 771 & 0.26 &&1.8& 2.1 &\\
\hline
\end{tabular}
\caption{\label{tab:fitpar}Parameters of the zero-range model ($a_{-}$ and $\eta$) and the finite-range model ($r$ and $\phi$) for $R=-55a_{0}$ from fitting to the experimental data of Ref.~\cite{Wacker2018}. The last row presents the average values.  The shown values of $\chi^2$ are normalized by the number of data points.}
\label{table:fitpa_zeroP}
\end{table}

{\it Loss coefficient.} At $\rho \to \infty$, any solution to Eq.~(\ref{eqn:schroedinger}) for $E>0$ can be written as a combination of incoming and outgoing waves
\begin{equation}
\label{eqn:inf_condition_GH}
    \begin{aligned}
    f\left(\rho \to \infty \right)\to He^{-i\sqrt{2}k\rho}+Ge^{i\sqrt{2}k\rho},
    \end{aligned}
\end{equation}
where $k=\sqrt{m E/\hbar^2}$. It is intuitively clear that information about losses must be contained in the ratio $|G/H|$. The WKB method of hidden crossing theory can be used to confirm this~\cite{Nielsen1999,Sorensen2013,DIncao2018}. Within this theory, the recombination coefficient for a given value of $k$ is written as
\begin{equation}
\label{eqn:alphak}
    \alpha_{k}\left(k\right)=36\left(2\pi\right)^{2}\sqrt{3}\frac{\hbar}{mk^{4}}\left(1-\left\lvert\frac{G}{H}\right\rvert^{2}\right).
\end{equation}
We show in~\cite{SM} that the ratio $G/H$ depends on a complex parameter $A$ that defines short-range three-body physics (not fixed by the effective range) via
 \begin{equation}
 f\left(\rho \to 0\right)\sim \sqrt{k\rho}\left(A+\ln{(\rho/|R|)}\right).
 \label{eq:boundary_A}
 \end{equation}
This parameter is determined by fitting to the experimental data, see Table~\ref{tab:fitpar} for typical values of $r$ and $\phi$ ($A=re^{i\phi}$). 

The recombination coefficient for a fixed temperature can now be obtained by thermally averaging with the Boltzmann distribution~\cite{Dincao2004,DIncao2018}
\begin{equation}
\label{eqn:alpha}
    \alpha=\frac{1}{2\left(k_{B}T\right)^{3}}\int\alpha_{k}E^{2}e^{-E/k_{B}T}dE.
\end{equation}
Here, we have assumed that a Bose gas forms a thermal cloud and that it is so dilute that many-body effects can be neglected. 
$\alpha$ from Eq.~(\ref{eqn:alpha}) fits well the experimental data. We illustrate this in Fig.~\ref{fig:datafit} using the data from Ref.~\cite{Wacker2018}. The figure shows the fit based upon the finite-range model from Eq.~(\ref{eqn:alpha}) together with the zero-range model (obtained with $R=0$, see also~\cite{Rem2013,Wacker2018}). Both fits describe the data equally well, i.e., they lead to similar values of ``chi-squared'',
see Table~\ref{table:fitpa_zeroP}.

\begin{figure}
\centering
\includegraphics[]{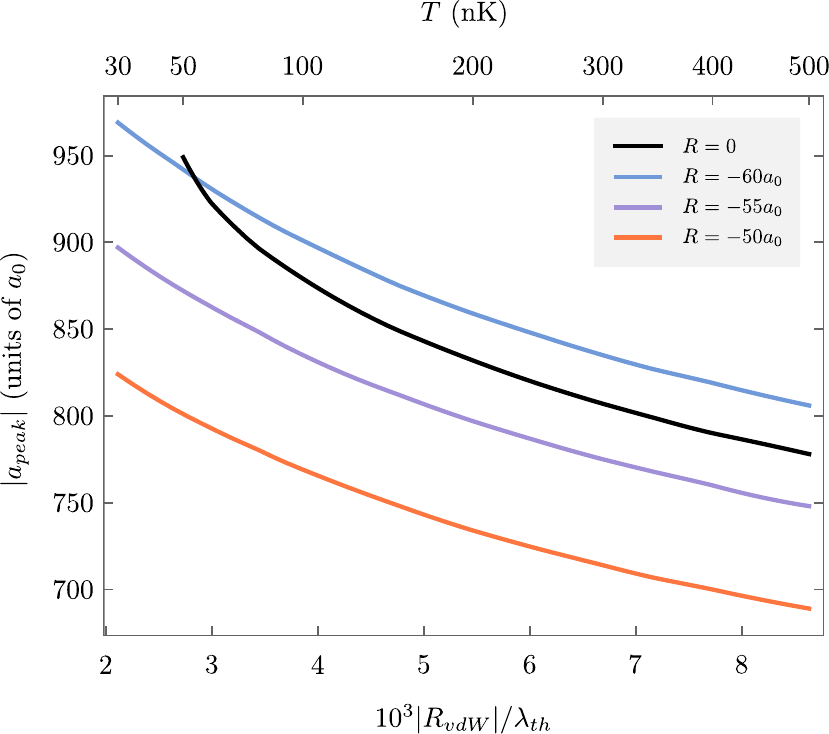}
\caption{\label{fig:peak_zero_P0}Peak of recombination, $|a_{peak}|$, as a function of temperature for different values of the effective range. $\lambda_{th}=\hbar/\sqrt{2\pi m k_{B}T}$~(cf.~\cite{PhysRevA.91.063622}), so $10^{3}|R_{vdW}|/\lambda_{th}\propto \sqrt{T}$. Parameters of the finite-range model: $r=1.8$, $\phi=2.1$; zero-range model has $|a_{-}|=771a_0$, $\eta=0.26$~(cf.~Table~\ref{table:fitpa_zeroP}).} 
\end{figure}

Equation~(\ref{eqn:alpha}) will be used in this work only to simulate `experimental' data, which are `realistic' in a sense that they contain beyond-zero-range effects.
However, note that the finite-range model can be used to understand beyond-zero-range physics in the context of finite-temperature effects. To motivate further analysis of the model, we present the temperature dependence of the recombination peak location $|a_{peak}|$ extracted from Eq.~(\ref{eqn:alpha}), see Fig.~\ref{fig:peak_zero_P0}. $|a_{peak}|$ increases for smaller temperatures, in agreement with previous studies, see, e.g.,~\cite{PhysRevLett.123.233402}.
This behavior is affected by the value of $R$. We observe that $|a_{peak}|(R_1)-|a_{peak}|(R_2)\sim 10(R_1-R_2)$ for the considered parameters. In-depth investigation of this scaling, which resembles  the van der Waals universality,
is left for future studies.

\begin{figure}[t]
\centering
\includegraphics[]{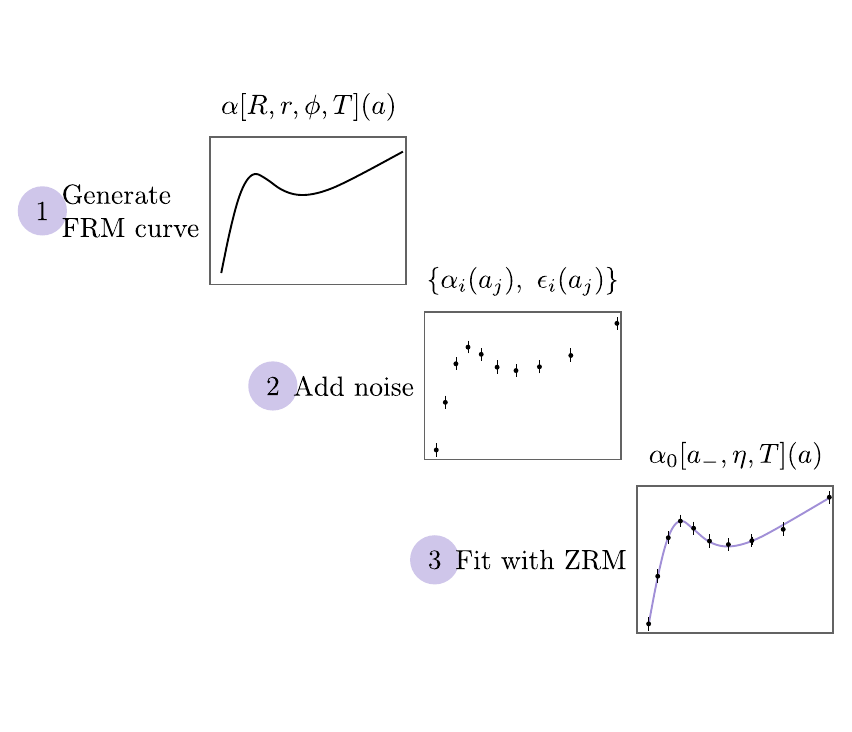}
\caption{\label{fig:model}Generation and analysis of `experimental' data. First, the recombination coefficient is calculated using the finite-range model (FRM). Second, this curve is used to generate `experimental' data points from a normal random distribution. The standard deviation is given by $\epsilon_i/2$, which is predetermined by us.
Third, the resulting artificial experimental data is fitted using the zero-range model (ZRM), which yields the parameter $a_{-}$.} 
\end{figure}

{\it Fitting the finite-range model.} Using the finite-range model, we generate `experimental' data for $^{39}$K, see a sketch in Fig.~\ref{fig:model}. As for the toy model, we consider here two `experiments' that measure $\{\{\alpha_i(a_1),\epsilon_i(a_1)\},...,\{\alpha_i(a_N),\epsilon_i(a_N)\}\}$ at different temperatures; $i=1$ ($i=2$) is for the first (second) `experiment'. For each $T$ and $a$, we draw values of $\alpha_i$ from a normal distribution whose mean is given by Eq.~(\ref{eqn:alpha}) with $r=1.8$ and $\phi=2.1$ (motivated by Table~\ref{tab:fitpar}). The standard deviation is given by $\epsilon_i/2$, which implies that the `experimental' (2-sigma) error bar is $\epsilon_i$. We assume that the two `experiments' measure at identical values of the scattering length (chosen in agreement with experimental points of Ref.~\cite{Wacker2018}), which can be determined exactly. The difference between the `experiments' is only in the values of $\epsilon_i$.  Similarly to the toy model, we work with $\epsilon_1(a)$, which is independent of $a$, and $\epsilon_2(a)$, which is proportional to $\alpha(a)$.

We use $\epsilon_1(a)=10^{-23}$ cm\textsuperscript{6}/s and $\epsilon_2(a)=\alpha(a)/20$. This implies that all points are equally trustworthy in the first `experiment', and the second `experiment' has the strongest confidence in the measurements in the non-universal region.

\begin{figure}[t]
\centering
\includegraphics{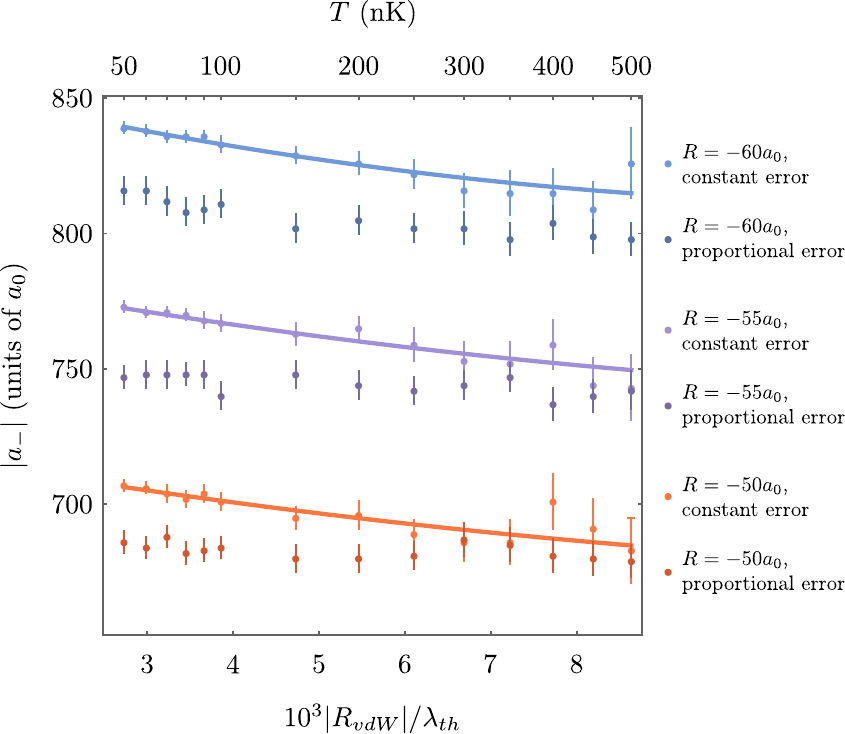}
\caption{\label{fig:genexpdata_zero_P0} The three-body parameter $a_{-}$ obtained from the artificial experimental data generated using the finite-range model with fixed parameters (see Fig.~\ref{fig:model}). Finite-range model parameters: $r=1.8,~ \phi=2.1$. Solid lines correspond to a direct zero-range model fit of the finite-range model (no added noise and error bars).} 
\end{figure}

The resulting data are fitted using the zero-range model~\cite{Braaten2008, Rem2013, Wacker2018} with the standard parametrization $a_-$ and $\eta$. 
The former parameter is shown in Fig.~\ref{fig:genexpdata_zero_P0} as a function of temperature for different values of effective range $R$. 

The figure shows that the extracted value of $a_{-}$ strongly depends on the experimental conditions~\footnote{Note that the change in the value of $a_-$ is similar to that of $a_{peak}$: $|a_-(R_1)|-|a_-(R_2)|\sim 10 (R_1-R_2)$.}. For the constant-error `experiment', there is a linear dependence of $a_-$ on $\sqrt{T}$ which agrees with~\cite{PhysRevA.91.063622,Wacker2018}, although with a different slope. The linear dependence is also seen in the direct fitting of the finite-range model with the zero-range model (without generation of `experimental' data).
In the proportional-error `experiment', we observe that $a_-$ is almost temperature-independent in agreement with~\cite{PhysRevLett.123.233402}. These results suggest that the difference between experimental observations of  Ref.~\cite{PhysRevLett.123.233402} and~Refs.~\cite{PhysRevA.91.063622,Wacker2018} might be explained by the difference in the experimental set-ups. 
Admittedly, other explanations cannot be ruled out at the moment. Experiments of~\cite{PhysRevA.91.063622,PhysRevLett.123.233402} might reach different conclusions because they focus on different systems (Cs vs K) and Feshbach resonances. The density of K cloud in Ref.~\cite{Wacker2018} was probably too high at low temperatures so that many-body effects could have played a role, see also a discussion in Ref.~\cite{PhysRevLett.123.233402}. 

In any case, the existing experimental data should be re-analyzed in light of our results. Indeed, the extraction of $a_-$ at $T=0$ from the data sets in Fig.~\ref{fig:genexpdata_zero_P0} leads to conflicting results implying that one needs additional information for identifying the `correct' universal value. The difference between the extracted values of $a_-(T=0)$ in the present example can be more than $5\%$, which is similar to the accuracy of the state-of-the-arts values~\cite{PhysRevLett.123.233402} and, thus, can be decisive in determining the error bars. This estimate suggests the following rule of thumb: a systematic error due to fitting with the zero-range model is of the order of $|R/a_-|$ (cf.~\cite{Braaten2008}).

Finally, we note that the sign of the slope of $a_{-}(T)$ in the first `experiment' (see Fig.~\ref{fig:genexpdata_zero_P0}) is determined by the sign of $R$, see~\cite{SM}.  This can be anticipated from the fact that the contribution of $R$ to the hyperspherical potential has a linear in $R$-term, which is perturbative, see Eq.~(\ref{eq:large_R_limit}).

{\it Summary and Outlook.}  
 We argued that the temperature dependence of three-body parameters may reflect certain characteristics of the experiment, and not the underlying three-body physics. In particular, it may reflect our confidence in accuracy of different data points.

We first considered a toy model in which the universal parameter (the slope of a line at $x\to\infty$) cannot be extracted reliably by considering only a finite range of $x$, no matter how many data points are produced by the `experiment' and how accurate they are. Most importantly, different distributions of error bars in the two considered `experiments' lead to different fitting parameters, i.e., the conclusions of these `experiments' are conflicting.

Then, we developed a finite-range model of three-body recombination and showed its good performance in describing experimental data. We used this model to simulate an `experiment' for a user of an (incomplete) zero-range model. As for the toy model, we showed that the type of error bars can change the value of the extracted fitting parameter $a_{-}$ by a few percent. This  leads to a systematic error in the value of the universal three-body parameters.

Our results might help to reconcile experimental observations of the dependence of $a_-$ on  $T$~\cite{PhysRevA.91.063622,PhysRevLett.123.233402, Wacker2018}. They may also motivate researchers to find approaches for extracting universal parameters from cold-atom data contaminated by non-universal physics. For three-body loss, the safest way is to provide measurements at larger scattering length. However, it is demanding from the experimental point of view. Alternatively, one can focus on available `true' observables, such as the peak position of  losses or estimate systematic error bars by assigning different weights to the points with smallest values of $|a|$~\footnote{The simplest way to implement the latter suggestion is to exclude a few points from the most non-universal region (e.g., around $x_0$ in the toy model) and evaluate the effect of this on the extracted parameter.  One must be careful when doing this in the analysis of three-body recombination as there should be enough points smaller than $a_-$ to ensure an accurate result. }. Finally, one can use theoretical finite-range models (such as the one presented above) to model experiments using standard Monte Carlo techniques~\cite{MC} and subsequently estimate possible systematic bias.

\section*{Acknowledgements}
We thank Jan Arlt, Hans-Werner Hammer, and Karsten Riisager for useful discussions.
M.L. acknowledges support by the European Research Council (ERC) Starting Grant No. 801770
(ANGULON).

\bibliography{td}

\end{document}